\documentclass[prb,preprint]{revtex4-2} 
\usepackage{amsmath}  % needed for \tfrac, \bmatrix, etc.
\usepackage{amsfonts} % needed for bold Greek, Fraktur, and blackboard bold
\usepackage{graphicx} % needed for figures
\usepackage{subcaption}
\begin{document}

% Be sure to use the \title, \author, \affiliation, and \abstract macros
% to format your title page.  Don't use lower-level macros to  manually
% adjust the fonts and centering.

\title{Experimental verification of the conservation of the magnetic moment and the longitudinal invariant}
% In a long title you can use \\ to force a line break at a certain location.

%When submitting the manuscript for review, do not include the author's name or institution
%\author{Daniel V. Schroeder}
%\email{dschroeder@weber.edu} % optional
%\altaffiliation[permanent address: ]{101 Main Street, Anytown, USA} % optional second address
% If there were a second author at the same address, we would put another 
% \author{} statement here.  Don't combine multiple authors in a single
% \author statement.
%\affiliation{Department of Physics, Weber State University, Ogden, UT 84408-2508}
% Please provide a full mailing address here.

\author{Juan Carlos Agurto}
\thanks{These authors contributed equally to this work.}
\affiliation{Departamento de F\'isica, Facultad de Ciencias, Universidad de Chile, Las Palmeras 3425, \~Nu\~noa, Santiago, Chile.}
\author{Felipe Darmazo}
\thanks{These authors contributed equally to this work.}
\affiliation{Departamento de F\'isica, Facultad de Ciencias, Universidad de Chile, Las Palmeras 3425, \~Nu\~noa, Santiago, Chile.}
\author{Amanda Guerra}
\thanks{These authors contributed equally to this work.}
\affiliation{Departamento de F\'isica, Facultad de Ciencias, Universidad de Chile, Las Palmeras 3425, \~Nu\~noa, Santiago, Chile.}\author{Erick Burgos-Parra}
%\email{ajp@dickinson.edu}
\affiliation{Departamento de F\'isica, Facultad de Ciencias, Universidad de Chile, Las Palmeras 3425, \~Nu\~noa, Santiago, Chile.}

% See the REVTeX documentation for more examples of author and affiliation lists.

\date{\today}

\begin{abstract}

Adiabatic invariants are fundamental to plasma physics but are often treated as purely theoretical concepts in undergraduate courses due to the difficulty of experimentally demonstrating them. This paper presents a pedagogical experiment to visualize and quantitatively verify the conservation of the magnetic moment ($\mu$) and the longitudinal invariant ($J$) using a standard educational electron charge-to-mass ratio apparatus configured as a magnetic bottle. By analyzing long-exposure photographs of the electron beam trajectory, we reconstructed the helical motion and calculated the invariants under different magnetic field configurations. Our results verify the conservation of the longitudinal invariant ($J$) with a ratio of 0.98 between configurations. The magnetic moment ($\mu$) exhibited a coefficient of variation of approximately 7\%, a deviation consistent with the presence of collisional effects in the tube. These findings demonstrate that complex plasma dynamics can be effectively studied using accessible laboratory equipment, providing a valuable bridge between theory and experiment for physics students.

% This study presents a pedagogical experiment designed to visualize and quantitatively validate two fundamental, yet often abstract, concepts in plasma physics: the adiabatic invariants. In a system known as a \textit{magnetic bottle}, we investigated the magnetic moment ($\mu$) and the longitudinal invariant ($J$). Using photographs of an electron beam emitted by an \textit{electron gun} of the PASCO Electron charge-to-mass ratio apparatus, that was placed inside the magnetic bottle, and subsequent data analysis, we obtained experimental values for $\mu$ and $J$ under two different configurations of current and potential. The average magnetic moment along the electron beam trajectory was $5.494 \times10^{-15}$ Nm/T with a coefficient of variation of 7.32\% in the first configuration, and $4.104 \times 10^{-15}$ Nm/T, with a coefficient of variation of $6.54\%$ in the second configuration. For the longitudinal invariant, we obtained values of $4.023 \,\times\,10^{5}$ m$^2$/s and $4.102 \,\times\, 10^{5}$ m$^2$/s for the first and second configurations respectively, corresponding to a ratio of $0.98$. Considering the experimental conditions and the methodology of the data analysis, we conclude that the conservation of $J$ was verified, while the results for $\mu$ demonstrate that $\mu$ is conserved within the limitations imposed by collisional effects and measurement uncertainty.
\end{abstract}
% AJP requires an abstract for all regular article submissions.
% Abstracts are optional for submissions to the "Notes and Discussions" section.

\maketitle % title page is now complete

\section{Introduction} % Section titles are automatically converted to all-caps.
% Section numbering is automatic.

An adiabatic invariant is a property of a system that remains constant under moderate\cite{chen} changes in its initial spatial and/or temporal conditions. Numerous studies have investigated both the conservation\cite{kaufmann} and the breakdown\cite{notte, haerendel} of the adiabatic invariants known as the magnetic moment and the longitudinal invariant. Most of these works, however, rely on satellite data or theoretical modeling, while direct experimental approaches often require setups that are difficult for undergraduate physics students to access. In this context, our work adopts a didactic perspective aiming to design an experiment that can be reproduced by students using equipment readily available in a standard physics laboratory. At the same time, our experiment can be regarded as a first step toward familiarizing students with experimental plasma physics\cite{pottersimeon}, a field in which research is typically associated with high financial costs and significant technical challenges.

\subsection{Magnetic Bottle}
\begin{figure}[h!]
    %\centering
    %\includegraphics[width=0.6\linewidth]{magnetic_bottle_schematic.png}
    % \includegraphics[width = 0.8\linewidth]{imagenes/botellamagnetica.png}
    % \caption{A schematic representation of a magnetic bottle, illustrating the fields lines, two coils, and the trapped particle (shown in red). In our experiment, instead of a single particle, we used an electron beam.}
    % \label{fig:esquemabotellamagnetica}
    % \begin{subfigure}{0.45\textwidth}
    %     \centering
    %     \includegraphics[width=\textwidth]{imagenes/helmholtz.png}
    %     \caption{Campo en bobina helmholtz}
    %     \label{fig:helmholtz}
    % \end{subfigure}

    % \hfill

    % \begin{subfigure}{0.45\textwidth}
    %     \centering
    %     \includegraphics[width=\textwidth]{imagenes/botellamagnetica.png}
    %     \caption{Botella magnetica}
    %     \label{fig:botellamagnetica}
    % \end{subfigure}

    % \begin{minipage}{0.45\textwidth}
    %     \centering
    %     \includegraphics[width=\textwidth]{imagenes/helmholtz.png}
    %     \caption{Uniform magnetic field in a Helmholtz coil. This is obtained if the separation of the coils is the same distance as the radius of the coils.}
    %     \label{fig:helmholtz}
    % \end{minipage}
    % \hfill
    % \begin{minipage}{0.45\textwidth}
    %     \centering
    %     \includegraphics[width=\textwidth]{imagenes/botellamagnetica.png}
    %     \caption{A schematic representation of the non-uniform magnetic field in a magnetic bottle, illustrating the field lines, two coils and the trapped particle (shown in red). The magnetic field is stronger in the regions near the coils, where the field lines get narrowed. }
    %     \label{fig:botellamagnetica}
    % \end{minipage}  

        \centering
        \includegraphics[width=0.7\textwidth]{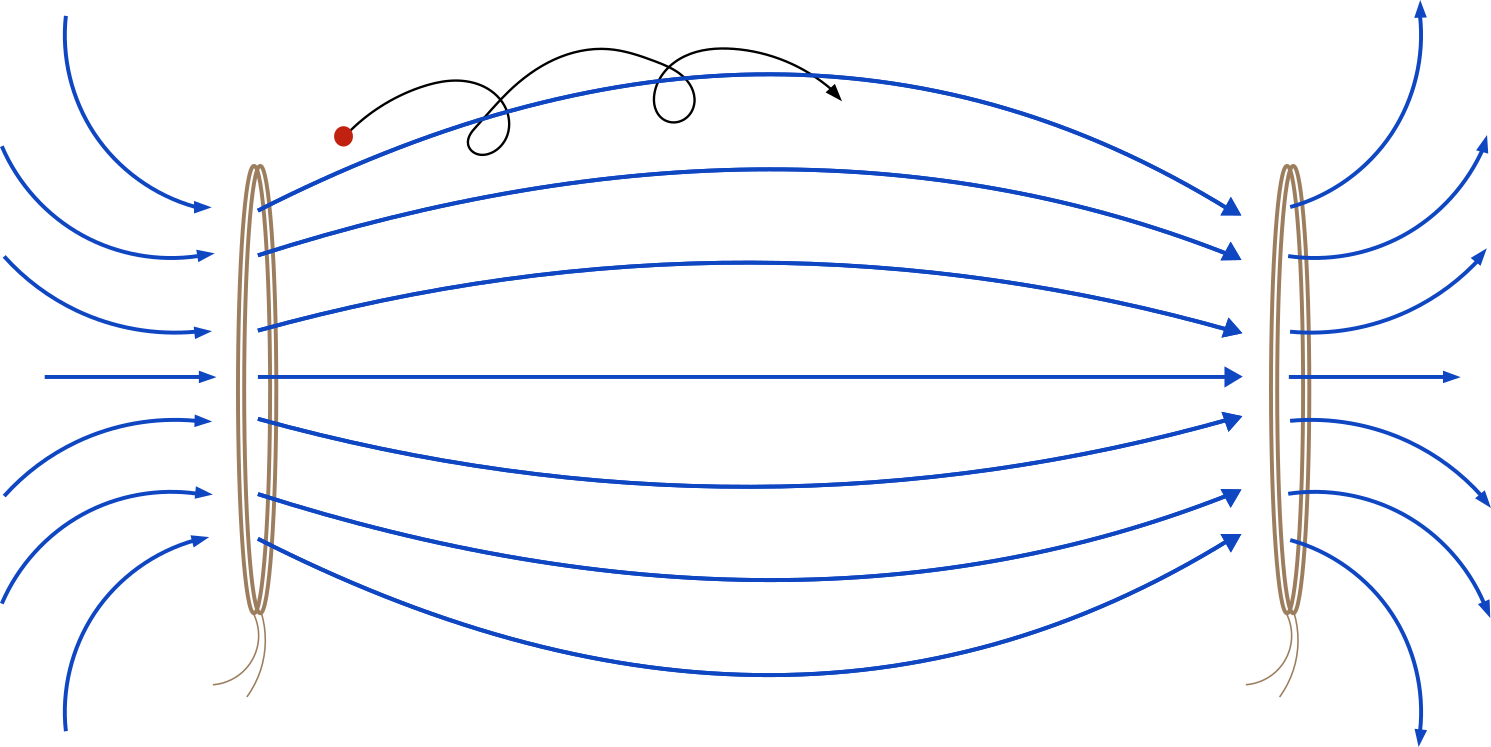}
        \caption{A schematic representation of the non-uniform magnetic field in a magnetic bottle, illustrating the field lines, two coils and the trapped particle (shown in red). The magnetic field is stronger in the regions near the coils, where the field lines get narrowed. }
        \label{fig:botellamagnetica}
    
\end{figure}

A magnetic bottle is a specific case of a magnetic mirror system, also referred to as a magnetic trap. Such systems confine charged particles within a defined region using magnetic fields\cite{grangerv, burdakov}. They generate magnetic fields with a gradient along the fields lines; thus, as a particle moves through a non-uniform magnetic field, its velocity component perpendicular to the field increases (due to conservation of the magnetic moment), while the parallel component decreases as it approaches the region of stronger magnetic field (the tighter side of the magnetic bottle, see Figure \ref{fig:botellamagnetica}). This propagation involves the conservation of energy, this means that when the particle reaches the \textit{mirror point} it will be reflected and reverses its motion along the same field line. This phenomenon is the focus of our study. %thus, as a particle propagates through the field, and considering energy conservation, its velocity component perpendicular to the field increases, while the parallel component decreases until it vanishes, causing the particle to bounce, a phenomenon that is the focus of this study.

\subsection{First adiabatic invariant: Magnetic moment}
The magnetic moment, $\mu$, is associated with the component of the particle's (electron) velocity perpendicular to the magnetic field\cite{chen, gurnett} and is responsible for the bounce effect in a magnetic bottle. It is defined as
\begin{equation}
    \mu \,=\, \dfrac{m \, v_{\perp}^2}{2 \,B} \,,
    \label{eq:mu}
\end{equation}
where $m$ is the mass of the trapped particle, $v_{\perp}$ is the velocity component perpendicular to the magnetic field, and $B$ is the magnetic field strength. According to theory, the magnetic moment remains constant under spatial variations of the magnetic field provided the condition $\omega \,<\,\omega_c$ is satisfied. Here, $\omega$ characterizes the rate of change of the magnetic field as experienced by the particle, while $\omega_c$ is the cyclotron frequency, given by 
\begin{equation}
    \omega_c \,=\, \dfrac{q\,B}{m}\,,
    \label{eq:omega_c}
\end{equation}
with $q$ denoting the charge of the trapped particle. This frequency correspond to the rate at which a charged particle gyrates around a magnetic field line. As a local property of the particle, and measurable through equation \ref{eq:mu}, the conservation of this invariant can be verified within a single configuration of initial conditions.

\subsection{Second adiabatic invariant: Longitudinal invariant}
The longitudinal invariant is associated with the component of a particle's velocity parallel to the magnetic field\cite{chen, gurnett}. It is defined as
\begin{equation}
    J \,=\, \int_{a}^{b} v_{\parallel}\,ds\,,
    \label{eq:jota}
\end{equation}
where the integration limits $a$ and $b$ denote the bounce points. This invariant ensures that a charged particle subjected to a magnetic field will always return to the same field line\cite{gurnett}. Conservation of the longitudinal invariant holds when temporal variations of the magnetic field are negligible compared to the particles's bounce period.
Consequently, two or more distinct configurations of the initial conditions are required to verify this conservation.

\subsection{Electron beam speed}
Since we are working with an electron beam rather than a single particle, it is necessary to determine its velocity at a given instant. Using the principle of energy conservation, the velocity can be calculated as
\begin{align}
    q\,V \,&=\, \dfrac{1}{2} \,m\,v^2\,\nonumber,\\
    v \,&=\, \sqrt{\dfrac{2\,q\,V}{m}}\,,
    \label{eq:e_vel}
\end{align}
where $v$ is the velocity, $q$ is the electron's charge, $m$ is its mass, and $V$ is the applied potential difference. Accurately determining this velocity is crucial, as it plays a key role in testing the conservation of the adiabatic invariants.
\newline
\newline
\indent The primary objective of this study was to experimentally test, through data acquisition and analysis, the conservation of the magnetic moment and the longitudinal invariant in a system known as a magnetic bottle.
%%%%%%%%%%%%%%%%%%% PROCEDIMIENTO %%%%%%%%%%%%%%%%%%%%
\section{Experimental Procedure}
\begin{figure}[h!]
\centering
\includegraphics[width=5in]{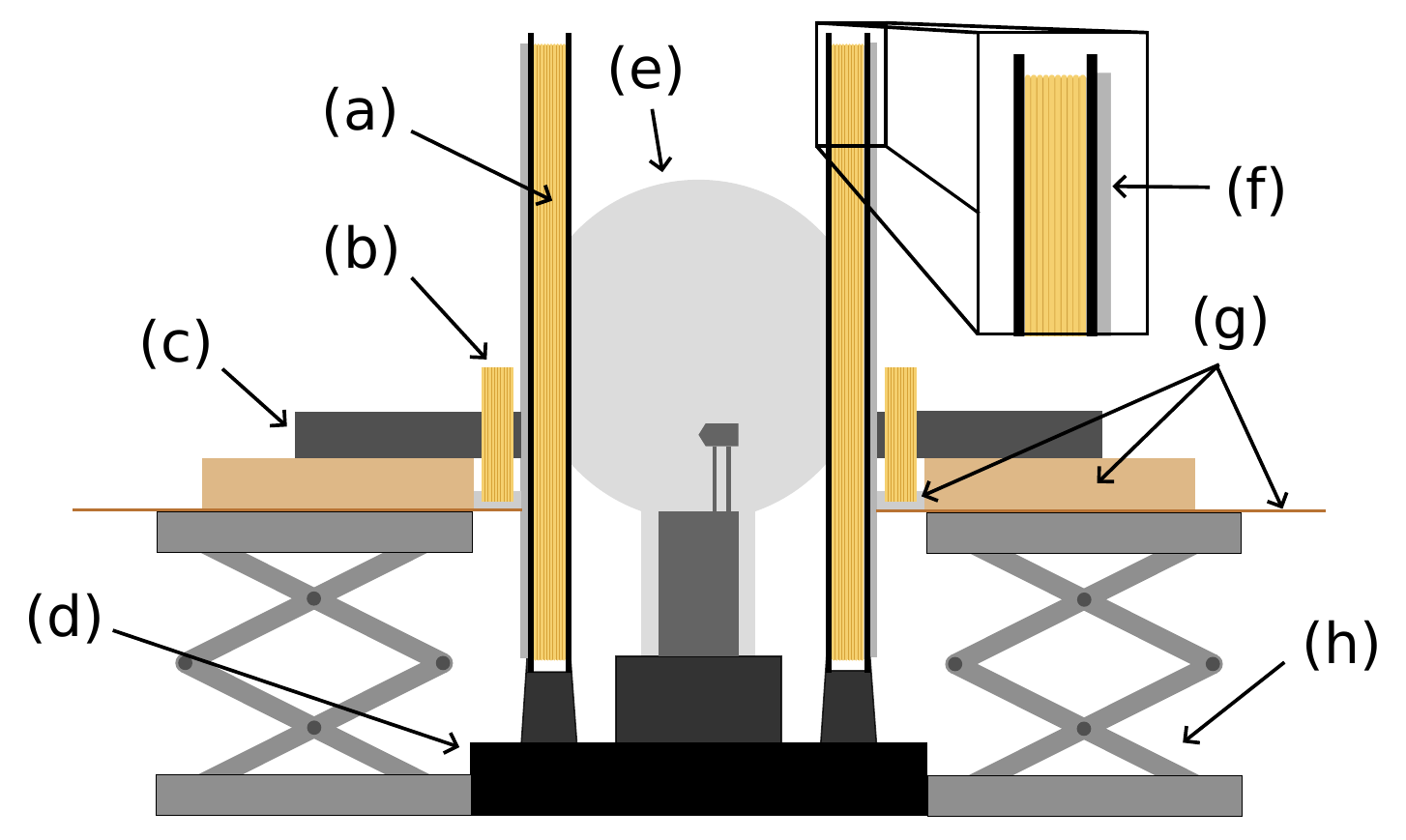}
\caption{The schematic of the experimental setup depicts: (a) the large coils (from PASCO SE-962), (b) the small coils (homemade), (c) the iron cores, (d) the base for the charge-to-mass setup (from PASCO SE-962), (e) the helium gas bulb, (f) the acrylic sheet, (g) miscellaneous supports for securing the small coils, and (h) a laboratory jack.}
\label{setup}
\end{figure}

To conduct the experiment, we employed a magnetic bottle and a helium-filled bulb containing an electron gun (shown in Figure \ref{setup}). The magnetic bottle consists of two iron cores with measured dimensions of $(2.0\pm 0.1)\times 10^{-2}$ m in width, $(3.0\pm 0.1)\times 10^{-2}$ m in height, and $(13.0\pm 0.1)\times 10^{-2}$ m in length, along with two pairs of coils available in the university’s teaching laboratory. The first pair corresponded to the coils from the PASCO Electron Charge-to-Mass Ratio apparatus (SE-9629), each with a nominal $130$ turns, an average diameter of $(15.8\pm 0.1)\times 10^{-2}$ m (calculated as the mean of the inner and outer diameters), and a separation of $(15.0\pm 0.1)\times 10^{-2}$ m, taking into account that part of the tube remained inside them. The second pair consisted of laboratory-built coils with $134$ turns, an average measured diameter of $(5.4\pm 0.1)\times 10^{-2}$ m, and a separation of $(19.6\pm 0.1)\times 10^{-2}$ m, which includes the $(0.8\pm 0.05)\times 10^{-2}$ m total thickness added by the coil supports and other intervening elements. At the center of the magnetic bottle configuration, we placed the e/m Tube (SE-9659), mounted on its base—both components belonging to the same electron charge-to-mass experiment setup. To protect the tube from possible displacements of the iron cores due to the magnetic field generated by the coils, two $(0.3\pm 0.05)\times 10^{-2}$ m-thick acrylic sheets were inserted between each large coil and its adjacent smaller coil.
%
%The coils were powered by DC supplies that allowed precise current control. 
The large coils were connected to a BroLight Tunable DC Power Supply (Constant Current) SE-9622 (hereafter Power Supply 1), while the smaller coils were powered by a Frederiksen Power Supply 0–24 V AC/DC 3630.00 (hereafter Power Supply 2). The tube received current from Power Supply 1, and the acceleration voltage was provided by a Tunable DC Power Supply II (Constant Voltage) SE-9644 (hereafter Power Supply 3). The current resolution for Power Supplies 1 and 2 was $0.01$ A, and the voltage resolution for Power Supplies 2 and 3 was $0.1$ V.
%Info: La fuente de poder 1 tiene una resolución de 0.01 A, la fuente de poder 2 tiene una resolución de 0.01 A y 0.1 V y la fuente de poder 3 tiene una resolución de 0.1 V 
%texto en párrafo: la resolución de la corriente en las fuentes de poder 1 y 2 era de 0.01 A y la resolución del voltaje en las fuentes 2 y 3 era de 0.1 V

Data were collected using a digital camera Nikon D3500 capable of long-exposure photography in dark conditions (i.e: 55 mm lens, ISO:200, F7.1, Manual focus with 6-8-10 seconds shutter speed), a leveling device, and two tripods, which were selected due to their suitable characteristics for optimal utilization in two distinct positions. The level ensured proper camera alignment on both tripods. The first tripod (hereafter Tripod 1) was positioned at $300.0 \pm 0.05$ cm from the side of the setup at a height such that the camera lens center aligned with the tip of the tube’s electron gun. The second tripod (hereafter Tripod 2) was positioned at a height of $102.00 \pm 0.05$ cm over the setup, again ensuring alignment between the lens center and the gun tip along the same axis. Two fluorescent rulers (shown in Figure \ref{beam_sections})— which allowed them to be observed in low-light conditions —were used to associate the positions of the top and side view photographs, aligning them with the edges of the larger coils.
%Para asociar las posiciones de las fotografías de la vista superior y lateral se utilizaron dos reglas fluorescentes-lo que permitía observarlas en la oscuridad-alineadas con los bordes de las bobinas grandes.

After completing the setup, we carried out the experimental procedure, which involved capturing two photographs—a top view and a side view—of the pattern formed by the electron beam trajectory under two different parameter configurations for Power Supplies 1, 2, and 3 (see Table \ref{configurations}). These configurations corresponded to two distinct magnetic field strengths, while maintaining the same beam injection angle in both cases. It should be noted that when transitioning from one view to the other—e.g., moving the camera from Tripod 1 to Tripod 2—the sources were kept on in every configuration. This measure prevented potential modifications to the beam due to the sensitivity of the adjustment knobs on the power sources used.
%Cabe destacar que al hacer la trasición de una vista a otra-i.e. mover la cámara del trípode 1 al trípode 2 por ejemplo-en cada configuración las fuentes se mantenían encendidas, evitando así posibles modificaciones en el haz dada la sensibilidad de las perillas de ajuste en las fuentes utilizadas. 
Data acquisition was based on these photographs, using a common spatial reference, corresponding to the aforementioned fluorescent rulers, to extract discrete points along the electron beam trajectory.
\begin{table}[h!]
\centering
\caption{Configurations of Parameters Used in the Power Supplies. Two initial set  of parameters (Current applied to coils and acceleration potential) yield two different configurations}
\begin{ruledtabular}
\begin{tabular}{l c c}
Parameter & Configuration 1 & Configuration 2 \\
\hline	% horizontal line to separate headings from data
Current in Larger Coils (A) & $3.15$ & $3.33$  \\
Current in Smaller Coils (A) & $0.63$ & $1.37$  \\
Acceleration Potential (V) & $138.0$ & $138.8$  \\
\end{tabular}
\end{ruledtabular}
\label{configurations}
\end{table}
\subsection{Data Extraction from Photographs}
Prior to the photographic acquisition process, it was crucial to ensure that the electron beam followed a helical trajectory along a straight line. This was necessary to ensure that the magnetic field was pointing in the $\hat{x}$ direction of the reference system used. 

During image acquisition, particular care was taken to avoid any rotation of the camera relative to the beam trajectory. In each image, one edge of the frame was aligned with the direction of the electron beam, depending on the beam’s orientation. The objective of this was to match our reference system with the coordinates used in the image processing, thereby avoiding the need to rotate the photographs.

\begin{figure}[h!]
\centering
\includegraphics[width=3in]{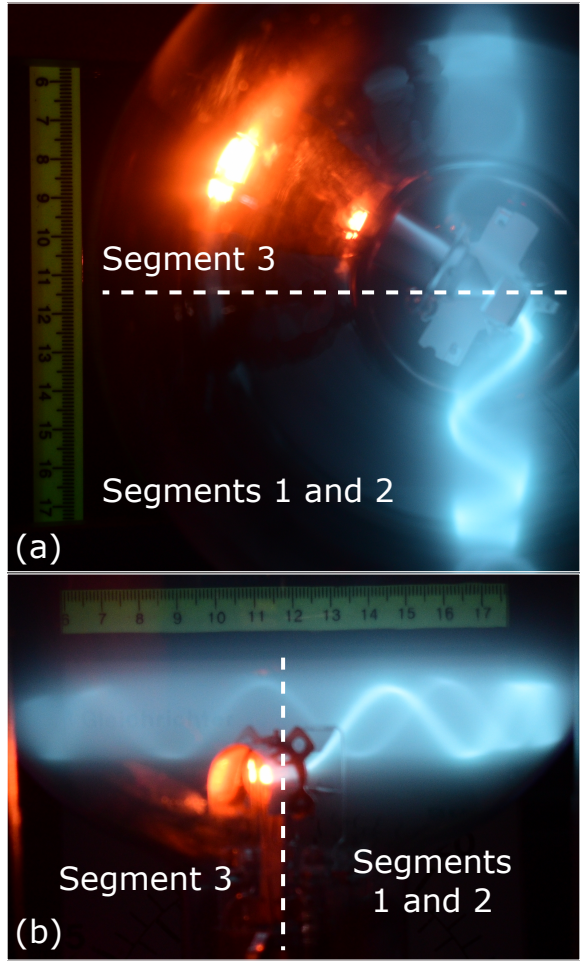}
\caption{Electron Beam Trajectory Divided into Segments. The figure shows (a) the top view and (b) the side view of the electron beam, where its trajectory was divided into three segments for observation. Segment 1 follows the trajectory from the trigger tip to the first bounce, Segment 2 follows the trajectory from the first bounce to the height of the trigger tip, and Segment 3 follows the trajectory from this latter point to the point of the second bounce.}
\label{beam_sections}
\end{figure}

After obtaining the photographs, the trajectory of the electron beam was analyzed by dividing it into three segments, each assigned to a different member of the team. These segments are shown in Figure \ref{beam_sections}.
For each parameter configuration, two photographs of the electron beam were taken — one from a top view and another from a side view. Data extraction was performed using the image software Inkscape\cite{Inkscape}. We began with the top view, establishing a common reference using the aligned fluorescent rulers visible in both image perspectives. The pixel coordinates of points along the beam were then recorded and converted into distances relative to this reference. These values were subsequently transformed into centimeters using a conversion factor obtained by measuring one centimeter on the ruler with the Measure Objects (M) tool in Inkscape.

Once the data from the top view was collected, we proceeded with the side view. The common axis between both perspectives was aligned using the reference point and the corresponding pixel-to-centimeter conversion factor for that image. This procedure allowed us to obtain side-view coordinates that directly corresponded to those extracted from the top view.
\subsection{Determination of the Magnetic Field}

%Falta poner el esquema del campo simulado en FEMM (mi compu me está dando problemas con la edición, pero la imagen ya está)
%
\begin{figure}[h!]
\centering
\includegraphics[width=5in]{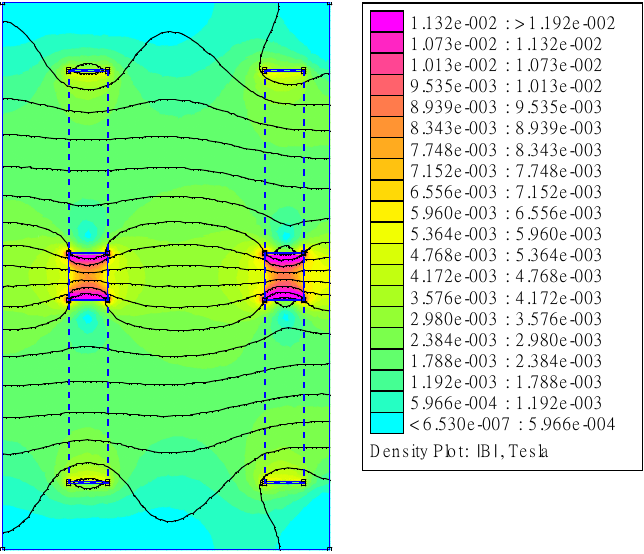}
%sin traducir: Esquema de la simulación realizada en FEMM del campo magnético de la configuración de bobinas. A la izquierda se puede observar el mapa de densidad del campo magnético, donde las líneas de campo se representan por las curvas de color negro. En la imagen las líneas azules continuas corresponden al contorno de las bobinas pequeñas, mientras que las líneas azules segmentadas corresponden al contorno de las bobinas grandes. A la derecha se encuentra la leyenda, la cual indica la correspondencia entre el orden del campo y los colores en la imagen de la izquierda.
%la traducción tiene correcciones con las que le pedí ayuda a gemini para ser más rigurosa con la terminología (siéntanse libres de corregir lo que estimen conveniente)
\caption{FEMM simulation scheme of the magnetic flux density for the coil configuration. The left panel shows the magnetic flux density plot, where the magnetic flux lines are represented by black curves. Solid blue lines correspond to the boundary of the small coils, while dashed blue lines indicate the boundary of the large coils. The right panel contains the legend (color scale), which correlates the magnetic flux density magnitude with the colors in the left image.}
\label{esquema_femm}
\end{figure}

To determine the magnetic field produced by the magnetic bottle in each configuration, we employed the software FEMM (Finite Element Method Magnetics)\cite{FEMM}, accessed via the Python library PyFEMM\cite{PyFEMM}. In the simulation, we assumed that the iron cores completely filled the interiors of the smaller coils and that  the acrylic sheets between the large and small coils weren’t present and, for simplicity of the simulation, that the two pairs of coils were concentric (see Figure \ref{esquema_femm}). 

%borrador (puede tener cambios la traducción): El campo magnético fue medido a lo largo del eje de rotación del haz de electrones comenzando desde el centro de la separación entre las bobinas grandes desplazándose en intervalos de 0.01 m desde ese punto hacia cada lado.
The magnetic field was measured along the axis of rotation of the electron beam, starting from the center of the gap between the large coils and extending outwards in $0.01$ m steps to both sides. The difference between the measured values and the values obtained from the FEMM simulation was approximately $0.5$ mT, with the magnetic field magnitude being on the order of milliTeslas.

%
%%%%%%%%%%%%%%%%%%%%%%%%%%%%%%%%%%%%%%%%%%%%%%%%%%%%%%
\section{Results}

\subsection{Data treatment}
The analysis focused on reconstructing the electron trajectories and determining their velocity components along each axis. The first step was to select reference points and record their pixel coordinates. This allowed defining a distance in centimeter for the $i-$th coordinate as
\begin{equation*}
    x_{i-\text{cm}} \,=\, \dfrac{|x_i \,-\, x_{i-\text{ref}}|}{P_x}\,,
\end{equation*}
where $x_{i-{\text{ref}}}$ is the reference point in pixels for the $i-$th coordinate, $x_i$ is the coordinate in pixels, and $P_x$ represent the number of pixels corresponding to $1$ cm in the analyzed image. The reference points were selected so that the origin of the reconstructed trajectories coincided with the origin used in the magnetic field simulations.

Once the coordinates were determined, the corresponding velocities were calculated. To do so, the differential element $ds$ was calculated using
\begin{equation*}
    ds \,=\, \sqrt{dx^2 \,+\, dy^2 \,+\, dz^2}\,,\,\,\, \text{con:} \,\,dx\,=\,x_{i+1}\,-\,x_i 
\end{equation*}
and the path parameter $s$ was obtained by doing the cumulative sum over $ds$ values.

\begin{figure}[h!]
    \centering
    \includegraphics[width=0.5\linewidth]{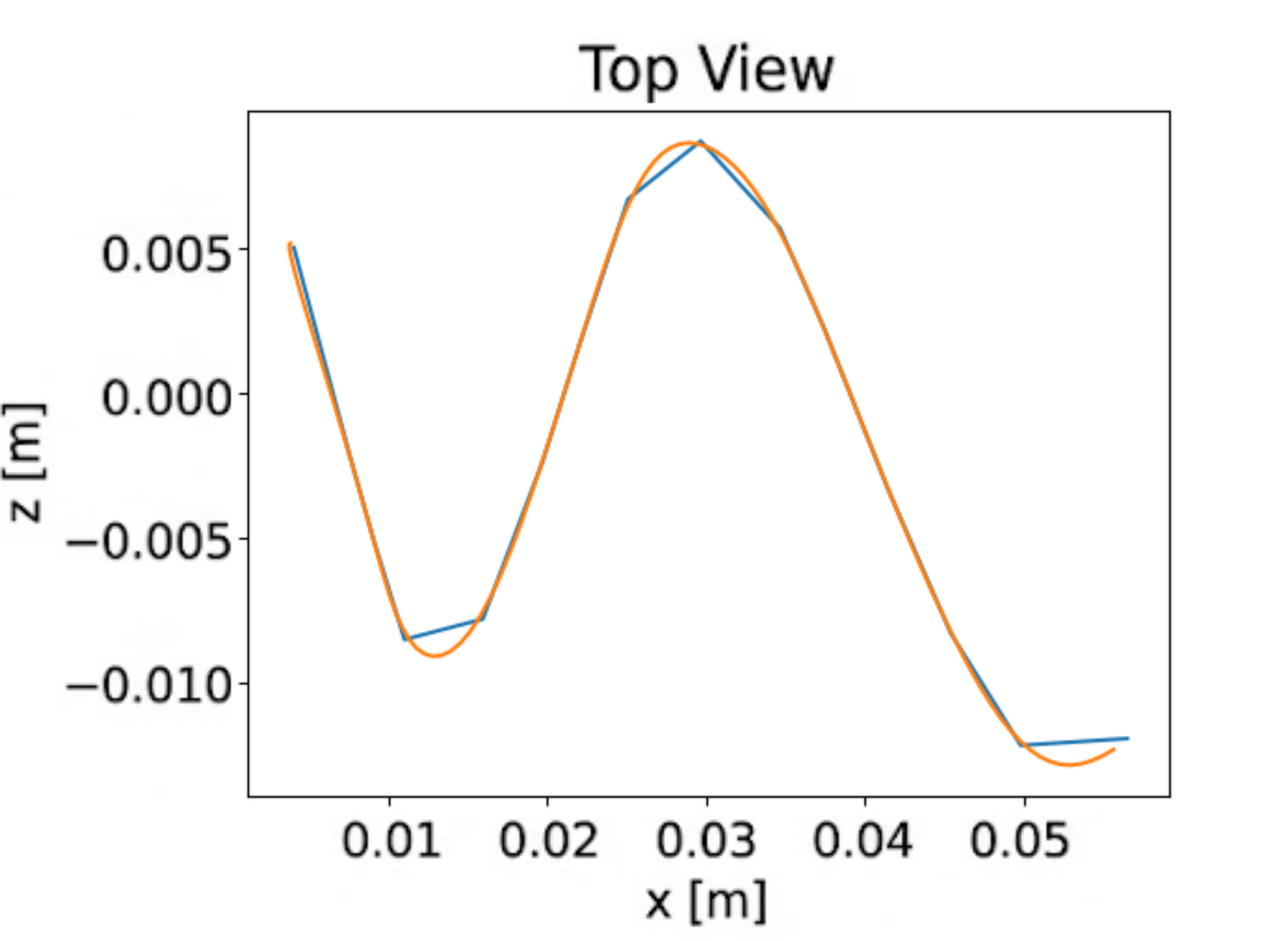}
    \caption{Electron beam trajectory graph for the top view of Segment $1$ in Configuration $1$. The blue curve represents the original data, meanwhile the orange curve represents the data after the interpolation and the Gaussian filter.}
    \label{fig:vista_superior}
\end{figure}

Each coordinate was then interpolated with respect to $s$, providing a denser and smoother representation of the trajectory. To further reduce noise, a Gaussian filter with $\sigma=2$ was applied. Figure \ref{fig:vista_superior} shows an example of the resulting smoothed trajectory for the top view of Segment $1$ in Configuration $1$.

Using these trajectories we computed the perpendicular and parallel velocities following the process detailed in the appendix section, which were then fitted to study the general tendency and reduce effects by residual noise. The details of this fitting are also presented in the appendix section.

\begin{figure}[h]
    \centering
    \includegraphics[width=1.0\linewidth]{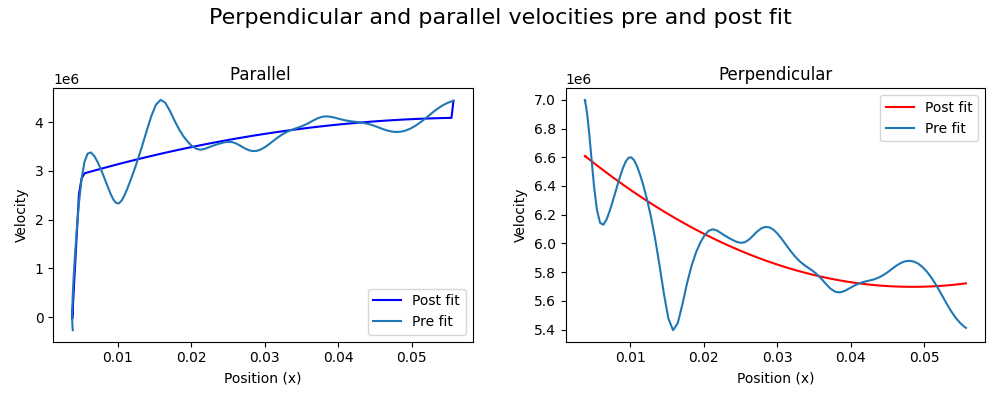}
 \caption{Calculated velocities for Segment 1 working under Configuration 1. Left side of the image shows the velocity's component parallel to the field , with the lighter blue being the velocity previous to the fit and darker blue after the fit. Right side the velocity's component perpendicular  to the field with light blue being the velocity previous to the fit and the red one after the fit}
   \label{velocidades pre y post}
\end{figure}

The results of this process are shown in Figure \ref{velocidades pre y post} where the velocities parallel and perpendicular to the magnetic field with their corresponding fit are presented. Using these velocities and equations \ref{eq:mu} and \ref{eq:jota} we calculate the adiabatic invariants and obtain the results presented in the following section


\subsection{Adiabatic invariants results}
As mentioned earlier, the first adiabatic invariant was evaluated separately for each magnetic field configuration and examined individually within each trajectory segment. The results are presented in Tables \ref{tab:config1} and \ref{tab:config2} for configurations $1$ and $2$ respectively. The reported value of $\mu$ corresponds to the mean of all calculated values along the trajectory, while CV represents the coefficient of variation, computed as the standard deviation divided by the mean, then multiplied by $100$.
\begin{table}[h!]
\centering
\caption{Results for $\mu$ in the Configuration $1$.}
\begin{ruledtabular}
\begin{tabular}{l c c}
Segment & $\mu$ $\left[\dfrac{\text{Nm}}{\text{T}}\right]$ & CV \\
\hline	% horizontal line to separate headings from data
1 & $5.514 \,\times\, 10^{-15}$ & 1.965\% \\ 
2 & $5.655 \,\times\, 10^{-15}$ & 7.315\% \\ 
3 & $5.314 \,\times\, 10^{-15}$ & 12.870\%   \\ 
Mean & $5.494 \,\times\, 10^{-15}$ & 7.323\% \\
\end{tabular}
\end{ruledtabular}
\label{tab:config1}
\end{table}
\begin{table}[h!]
\centering
\caption{Results for $\mu$ in the Configuration $2$.}
\begin{ruledtabular}
\begin{tabular}{l c c}
Segment & $\mu$ $\left[\dfrac{\text{Nm}}{\text{T}}\right]$ & CV \\
\hline	% horizontal line to separate headings from data
1 & $4.087 \,\times\, 10^{-15}$ & 2.897\% \\ 
2 & $4.219 \,\times\, 10^{-15}$ & 5.379\% \\ 
3 & $4.007 \,\times\, 10^{-15}$ & 11.344\% \\ 
Mean & $4.104 \,\times\, 10^{-15}$ & 6.540\% \\ 
\end{tabular}
\end{ruledtabular}
\label{tab:config2}
\end{table}
From these results, it is evident that $\mu$ does not appear to remain strictly conserved, as the average coefficient of variation exceeds the expected threshold of $5\%$ for conservation. This deviation can be attributed to two main factors: one physical and one methodological.

The theoretical factor arises from  the assumptions of the model. By considering total velocity conservation, the model implicitly neglects collisions. In practice, collisions occur both among electrons and between electrons and the background gas, breaking the cyclotron motion phase, affecting $v_{\perp}$ immediately. This directly affect the conservation of $\mu$\cite{chen}.

The methodological factor concerns the image acquisition process. The photographs used for analysis are subject to parallax and optical aberrations, which -even after careful correction- introduce measurement errors in position and distance, as well as variations in the pixel-to-centimeter conversion factor across different regions of the images. This issue is most pronounced in segment $3$, where this spatial scaling fluctuates significantly depending on where it is measured.

A closer look reveals that the results degrade progressively from segment $1$ to segment $3$. This trend supports the idea that collisions are the dominant source of error: as the electrons travel farther, they undergo a greater number of collisions, naturally increasing the coefficient of variation.

For the second adiabatic invariant ($J$), data from segments $2$ and $3$ were used, as these span the full motion from one mirror point to the other. The calculated values for each configuration were
\begin{align*}
    J_1 \,&=\, 402,285.569 \,\left[ \dfrac{\text{m}^2}{\text{s}} \right]\,,\\
    J_2 \,&=\, 410,191.790 \,\left[ \dfrac{\text{m}^2}{\text{s}} \right]\,.
\end{align*}
The ratio between these two values ($J_1/J_2$) is $0.98$, which strongly indicates that this invariant was conserved within experimental uncertainty.

Although this may seem surprising, given that segment $3$ was identified as the most error-prone, the result is physically reasonable. The longitudinal invariant represents a more global property of the motion; therefore it is less sensitive to local perturbations such as collisions. As long as the magnetic mirror effect occurs, the longitudinal invariant $J$ remains conserved.

\section{Conclusion}
In this work we experimentally tested the conservation of the magnetic moment and the longitudinal invariant of a system comprising a electron beam under a magnetic field in a magnetic bottle configuration.. To verify the conservation of the magnetic moment, we calculated the average value of $\mu$ along the electron beam trajectory for both magnetic configurations. The average obtained  for the Configuration 1 was $5.494 \times 10^{-15}$\,Nm/T with a coefficient of variation of $7.323\%$, while for the Configuration 2  the mean value was $4.104 \times 10^{-15}$\,Nm/T with a coefficient of variation of $6.54$\,\%. To assess the conservation of the longitudinal invariant, we computed its value in both configurations, obtaining $J_1$ for Configuration 1 and $J_2$ for Configuration 2. The ratio between these values is $0.98$, indicating that this quantity is preserved %poner si se conserva o no

%The results could have been improved by correcting for several systematic errors associated with the photographic capture process and additional considerations in the theoretical model. As discussed in the previous section the main errors associated with the first invariant arise from the parallax in the pictures and the collisions in the system that differs from the theoretical model
Overall, these findings demonstrate that the adiabatic invariants can be effectively characterized in our system, showing that with a relatively simple and acquirable system it is possible to study these quantities with satisfying results.  Considering  the experimental setup limitations and the analysis discussed previously, we can conclude that the results for the magnetic moment fall within a reasonable margin of expectation. On average, the values are close to the range consistent with conservation, and the segment-by-segment analysis shows behavior consistent with the presence of collisions. As for the second invariant $J$, its ratio between configurations indicates that this magnitude is close to being the same between configurations,which leads to conclude that it is indeed being conserved and that the observed discrepancy may arise from systematical errors such as the ones discussed before.%aquí iba lo de decir qué implicaba le proporción utilizada
Finally, beyond the numerical results, this experiment has relevance for the undergraduate physics curriculum. Typically, adiabatic invariants are presented to students only as mathematical derivations. This setup allows them to visualize the magnetic bottle effect and observe the difference between the cyclotron motion and the bounce motion.

Furthermore, the analysis method introduces students to necessary experimental techniques, such as image processing to extract data and the numerical integration of real signals. The use of FEMM simulations also helps to understand the connection between theoretical modeling and experimental measurements. Lastly, the deviations found in the magnetic moment results serve as a practical example to discuss the limitations of ideal models and the role of collisions in plasma systems. These aspects show that it is possible to study complex plasma concepts using accessible laboratory equipment. %aquí se hablaba del éxito en la comprobación de la conservación, falta corregir cosas en caso de que la invariante lu¿ongitudinal no haya arrojado un buen resultado en cuanto a su consrvación 

\appendix*   % Omit the * if there's more than one appendix.

\section{} %título del manuscrito de ejemplo, por cambiar

\subsection{Velocity calculation}
Having expressed each coordinate as a function of $s$: $x(s)$, $y(s)$, and $z(s)$, the time derivatives were computed using the chain rule as follows
\begin{equation*}
    v_{x_i} \,=\, \dfrac{dx_i}{dt} \,=\, \dfrac{dx_i}{ds}\,\dfrac{ds}{dt}.
\end{equation*}

Because the total kinetic energy is conserved, the total velocity $v$ remains constant, and $\dfrac{ds}{dt} \,=\, v$, therefore
\begin{equation*}
    v_{x_i} \,=\, \dfrac{dx_i}{ds}\,v
\end{equation*}

However, each of these steps contains numerical errors that accumulate and amplify in the following step. To correct for this, each velocity component was divided by the magnitude of the velocity vector, obtaining a unit velocity vector that, when multiplied by the known $v$, yielded the corrected velocity components. It is important to emphasize that the magnitude used for normalization was not the $v$ calculated from equation \ref{eq:e_vel}, but rather the one directly obtained from the experimental data, which ensures that the direction of the unitary vector is determined by the experimental data. Since the coordinate system was aligned so that the magnetic field was exclusively along the $x-$axis, the electron's velocity component parallel to the magnetic field line was given by 
\begin{equation*}
    v_{\parallel} \,=\, v_x\,,
\end{equation*}
where $v_x$ is the electron's velocity component along the $x-$axis. The perpendicular component was calculated as
\begin{equation*}
    v_{\perp} \,=\, \sqrt{v_{y}^{2} \,+\, v_{z}^{2}}\,,
\end{equation*}
where $v_y$ and $v_z$ are the velocity components along the $y$ and $z$ axes, respectively. Consequently, the total velocity satisfies
\begin{equation}
    v^2 \,=\, v_{\parallel}^{2} \,+\, v_{\perp}^{2}\,.
    \label{eq:vel_total}
\end{equation}

\subsection{Fitting the velocities}

To minimize residual noise and highlight overall behavior, the data was fitted, and the resulting fit was used as the final estimate of the velocity components.

\begin{figure}[h!]
    \centering
    \includegraphics[width=0.5\linewidth]{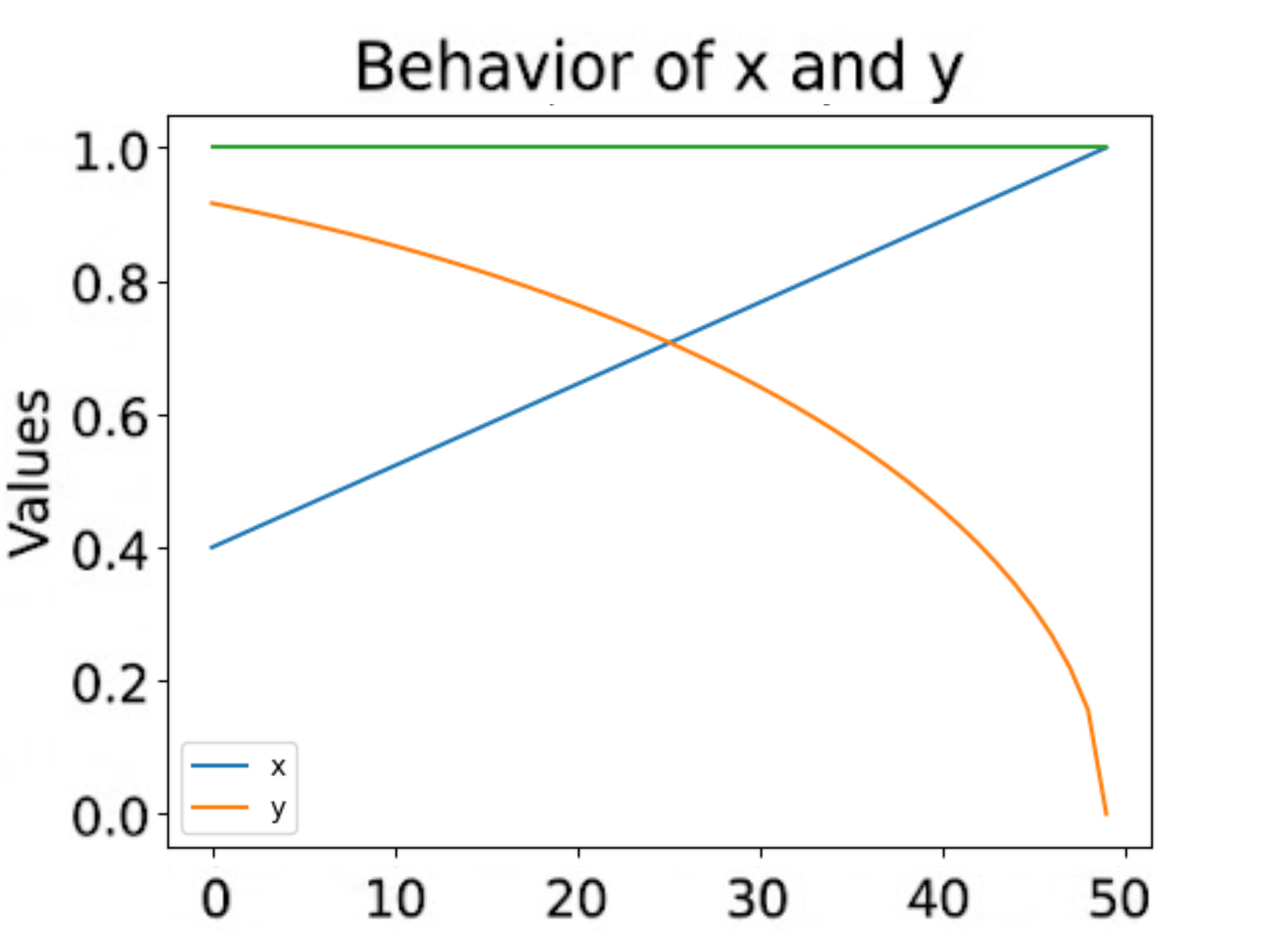}
    \caption{Plot of the behavior of arbitrary data $x$ and $y$ such that $x^2 \,+\, y^2 \,=\,1$.}
    \label{fig:compensatory}
\end{figure}

For the parallel velocity, the data were divided into two segments, as they exhibited significantly different trends. This difference arises from the relationship expressed in equation (\ref{eq:vel_total}), to compensate for the increase in perpendicular velocity near the mirror points, the parallel velocity must decrease more sharply. An example of this compensatory behavior is shown in Figure \ref{fig:compensatory}, where an arbitrary dataset $(x,\,y)$ satisfying the condition $x^2 \,+\, y^2 \,=\, 1$ is plotted. It can be seen that while $x$ increases at a constant rate, $y$ must decrease increasingly steeply to maintain the relationship.

With this in mind, and after a process of trial and error in which we computed both adiabatic invariants and the average percentage difference between the pre and post fit velocity data, we determined that the most accurate approach was to fit the $10\%$ of the data points closest to the mirror point using an exponential function, while fitting the remaining portion with a quadratic polynomial. The perpendicular velocity did not exhibit the same irregular behavior, therefore, the $100\%$ of its data were fitted with a quadratic polynomial. Figure  \ref{velocidades pre y post} show the velocity profiles before and after fitting. The average difference between the fitted and original data was approximately $10\%$ for the perpendicular velocity, and $15\%$ for the parallel velocity. It is important to note that this percentage difference increases across the segments shown in figure \ref{beam_sections}, ranging from about $3\%$ in the first segment to roughly $29\%$ in the third.
%\begin{figure}[h!]
%    \centering
%    \includegraphics[width=0.5\linewidth]{imagenes/perpendicular.png}
%    \caption{Plot of the perpendicular velocity before and after the fitting. In blue is shown the original data, in orange the fitted velocity.}
%    \label{fig:vel_perp}
%\end{figure}
%\begin{figure}[h!]
%    \centering
%    \includegraphics[width=0.5\linewidth]{imagenes/paralela.png}
%    \caption{Plot of the parallel velocity before and after the fitting. In blue is shown the original data, in red the fitted velocity.}
%    \label{fig:vel_parallel}
%\end{figure}

%\begin{acknowledgments}

%We acknowledge the 

%\end{acknowledgments}

\end{document}